\documentclass[sigconf]{acmart}

\usepackage[export]{adjustbox}
\usepackage{balance} 
\usepackage{booktabs, multirow, makecell}
\usepackage[multiple,para]{footmisc} 
\usepackage{graphics} 
\usepackage{microtype} 
\usepackage{subfig}
\usepackage[flushleft]{threeparttable}
\usepackage{booktabs}
\usepackage{dblfloatfix}
\usepackage{threeparttable}
\usepackage{siunitx}

\copyrightyear{2018}
\acmYear{2018}
\setcopyright{acmcopyright}
\acmConference[DH'18]{2018 International Digital Health Conference}{April 23--26, 2018}{Lyon, France}
\acmBooktitle{DH'18: 2018 International Digital Health Conference, April 23--26, 2018, Lyon, France}
\acmPrice{15.00}
\acmDOI{10.1145/3194658.3194663}
\acmISBN{978-1-4503-6493-5/18/04}

\begin{document}
\title{Estimating Glycemic Impact of Cooking Recipes via \\Online Crowdsourcing and Machine Learning}

\author{Helena Lee}
\affiliation{%
  \institution{Singapore Management University}
}
\email{helenalee@smu.edu.sg}

\author{Palakorn Achananuparp}
\affiliation{%
  \institution{Singapore Management University}
}
\email{palakorna@smu.edu.sg}

\author{Yue Liu}
\affiliation{%
  \institution{Singapore Management University}
}
\email{yueliu@smu.edu.sg}

\author{Ee-Peng Lim}
\affiliation{%
  \institution{Singapore Management University}
}
\email{eplim@smu.edu.sg}

\author{Lav R. Varshney}
\affiliation{%
  \institution{University of Illinois at Urbana-Champaign}
}
\email{varshney@illinois.edu}

\renewcommand{\shortauthors}{Lee et al.}

\begin{abstract}
Consumption of diets with low glycemic impact is highly recommended for diabetics and pre-diabetics as it helps maintain their blood glucose levels. 
However, laboratory analysis of dietary glycemic potency is time-consuming and expensive. In this paper, we explore a data-driven approach utilizing online crowdsourcing and machine learning to estimate the glycemic impact of cooking recipes. We show that a commonly used healthiness metric may not always be effective in determining recipes suitable for diabetics, thus emphasizing the importance of the glycemic-impact estimation task. Our best classification model, trained on nutritional and crowdsourced data obtained from Amazon Mechanical Turk (AMT), can accurately identify recipes which are unhealthful for diabetics.

\end{abstract}

%
%

\copyrightyear{2019} 
\acmYear{2019} 
\acmConference[DPH' 19]{9th International Digital Public Health Conference (2019)}{November 20--23, 2019}{Marseille, France}
\acmBooktitle{9th International Digital Public Health Conference (2019) (DPH' 19), November 20--23, 2019, Marseille, France}
\acmPrice{15.00}
\acmDOI{10.1145/3357729.3357748}
\acmISBN{978-1-4503-7208-4/19/11}

\ccsdesc[500]{Information systems~Information retrieval}
\ccsdesc[500]{Computing methodologies~Natural language processing}
\ccsdesc[300]{Applied computing~Consumer health}

\keywords{Glycemic Impact; Recipe Embeddings; Recipe Classification}

\maketitle

\section{Introduction}

Findings from several health studies have established a strong association between type-2 diabetes and diets \cite{Livesey2008}. People with diabetes (henceforth \textit{diabetics}) and those with high-risk factors (henceforth \textit{pre-diabetics}), such as overweight and obese groups, are recommended to consider dietary glycemic potency/impact --- the effect of diets on blood sugar levels after consumption (also known as \textit{glycemic response}) --- when making food choices to improve glycemic control. While the standardized \textit{in vivo} laboratory testing method is still the most precise way of measuring glycemic index (GI) of foods and beverages, it can be relatively costly and time-consuming \cite{Atkinson2008}. 
A naive approach of looking up the GI of items in GI databases is not practical. Many GI databases are not likely to cover food items in long-tail consumption, e.g., the published international tables of GI \cite{Atkinson2008} only contain approximately 2.5K records of common food items. In comparison, over one million cooking recipes can be found online \cite{Salvador2017}. Besides, the GI of a recipe cannot simply be derived by linearly combining the carbohydrate contents of its ingredients. Apart from GI, diabetics and pre-diabetics may consider dietary healthiness scores \cite{Trattner2017b}, which can be directly computed from the nutritional information of food items, thus more easily obtainable. However, the relationship between the glycemic potency of diets and their healthiness scores is not yet well understood.

We explore glycemic impact inference as a binary classification problem in which the goal is to classify whether a given recipe is \textit{unhealthy for diabetics} (UD) or \textit{healthy for diabetics} (HD) using textual and nutritional contents of the recipe. UD recipes are defined as those likely to cause elevated glycemic responses, i.e., having moderate to high glycemic impact, and vice versa. 
We utilized the knowledge and perception of AMT workers to obtain the glycemic impact label of recipes. Next, we conducted a comprehensive evaluation of various word and sentence embedding models to investigate their performance on UD recipe classification. Then, we analyzed the importance of recipe features, including textual and nutritional features, and the task performance. Our study seeks to answer the following research questions (RQs): \textit{What is the relationship between glycemic impact and healthiness scores of cooking recipes} (\textbf{RQ1})?; \textit{what is the effectiveness of different recipe representations on the UD recipe classification task} (\textbf{RQ2})?

\section{Related Work}

Past research has explored a data-driven approach to quantify the healthiness and predict the nutritional information of cooking recipes  \cite{Kusmierczyk2016,Rokicki2018}.
While the dietary glycemic impact cannot be directly derived from the recipes' ingredients or nutritional information \cite{Atkinson2008}, it is novel and interesting to automatically detect cooking recipes that are unhealthy for diabetics to consume.

The recipe representation problem has been recently investigated in cross-modal recipe retrieval \cite{Salvador2017,Wang2019} and recipe transformation \cite{kazama2018neural}. In contrast to prior work, we compare the performance of various recipe representation methods, including word, sentence, and paragraph embeddings, on the recipe classification task.

\section{Dataset} \label{data}

\subsection{Recipes Collection} \label{collect}

We first crawled 55,102 recipes from a popular online recipe website \textit{Allrecipes} by traversing its recipe categories directory. The crawling was performed from January 2019 to February 2019. Each recipe page consists of the following features: title, ingredients, cooking directions, and nutritional properties. Furthermore, we derived a \textit{dry weight} of each recipe by summing up the weight of all nutrition properties. The value of each property was then normalized by its dry weight. Together with dry weight, there are 20 nutritional properties per recipe. We discarded 1,026 incomplete recipes having fewer than two ingredients or two sentences in cooking directions. This dataset is designated \textbf{Recipe54K}. 

\subsection{Human Annotation of Glycemic Impact} \label{annotate}

\textbf{Data Selection}
We defined two criteria for selecting a subset of 1,000 recipes for the annotation tasks. First, the recipes should be representative of all levels of glycemic impact from low to high. Second, the recipes should be fairly difficult for multiple models to classify, thus requiring human judgements.
To satisfy the first criterion, we check the \textit{sugar-to-fiber} (S/F) ratio of recipes because this ratio could be considered as a distant proxy of glycemic impact as both sugar and dietary fiber are generally correlated with the dietary glycemic potency. \cite{Atkinson2008}. For the second criterion, we build classification models with two noisy sources of glycemic impact label: One is the \textit{low glycemic impact} category assigned to some recipes by Allrecipes; another is derived from the S/F ratio where the cut-off is heuristically set to 1. Then, we identified the most misclassified recipes based on the prediction probability. Lastly, we sampled recipes uniformly from the three S/F partitions with ratio cut-offs at 1 and 13. For each partition, we selected approximately 333 recipes with the lowest rank-sums of prediction probability. After these procedures, 1,000 recipes were selected and uploaded to AMT.

\textbf{Qualified Worker Selection} 
To select qualified workers with sufficient knowledge, we created a qualification task composed of six multiple-choice questions with known answers in AMT. One asked for the definition of glycemic impact. Three of the questions required the workers to select from a few ingredients the one with the highest or the lowest glycemic impact. The other two questions were about judging the glycemic impact of two cooking recipes. The design of these two questions was the same as the actual HITs described in the next section. Workers had to correctly answer 4 of 6 questions to qualify for our HITs. Ultimately, 20 AMT workers were qualified (48.7\% of those who ever contributed).

\textbf{Glycemic Impact-Based HITs}
For each HIT, we presented textual description of a recipe, including title, ingredients, and cooking directions, to a crowd worker and asked him/her to judge the glycemic impact of the given recipe according to the question "\textit{Do you agree this recipe has a low glycemic impact?}" The worker specified his/her response in a 5-point Likert scale from strongly agree (1) to strongly disagree (5). Nutritional information of the recipe was not shown to the worker. In addition, the workers were allowed to skip any questions if they were not confident in their judgement by choosing a "not sure" option. We encouraged the workers to use any online resources about relevant topics, such as diabetes, glycemic index, etc., to assist in their judgements. For each HIT, the workers were paid 0.10 USD as compensation. In total, three judgements were required per recipe. In the end, we collected 3,000 total judgments from 8 unique workers.

\textbf{Quality of Annotation}
Next, we measured the agreement between crowd workers using Krippendorff's alpha ($\alpha$) \cite{Krippendorff2011}. The value of $\alpha$ computed from the 3,000 judgments was 0.467, indicating a highly subjective nature of the task. We observed that many recipes containing ingredients with high-carbohydrate content but relatively moderate to low GI (e.g. pasta, black beans) were judged to have high glycemic impact, receiving a rating of 3 or lower (neutral to strongly disagree) from the workers. This may indicate precaution or uncertainty in their evaluations. Next, recipes with a rating of 4 or higher contained little to no carbohydrate content (causing no impact in glycemic response). Lastly, AMT workers disagreed the most when judging recipes for cocktail drinks, whole wheat dishes, and low-carb cheesecakes.

\textbf{Crowdsourcing Aggregation} 
We used Dawid-Skene algorithm \cite{Dawid1979} to determine the ground-truth labels from the 3,000 crowdsourced judgements. As a result, 990 recipes were assigned proper numerical labels ranging from strongly disagree (1) to strongly agree (5), whereas 10 recipes were assigned a "not sure" label. The dataset of 990 labeled recipes obtained through online crowdsourcing is designated \textbf{Recipe990}.

Then, we derived the positive and negative classes for the binary classification task to identify unhealthy-for-diabetics recipes. Based on the patterns of crowdsourced judgements, we considered any recipes with a rating of 3 or lower as positive examples (\textbf{UD}: unhealthy for diabetics) for the classification task. Conversely, those with a rating of 4 or higher were treated as negative examples (\textbf{HD}: healthy for diabetics). In total, there were 506 recipes in the positive class and 484 recipes in the negative class. 

According to aggregated statistics, UD recipes generally have higher carbohydrate and sugar contents than HD recipes, whereas HD recipes tend to have higher protein and fat contents. These observations seem to be consistent with the judgements of the AMT workers discussed earlier. Specifically, when evaluating the glycemic impact of recipes, the workers tended to strictly focus on ingredients relatively high in carbohydrates and total sugars even though they were not shown the recipes' nutrition information.

\subsection*{RQ1: Glycemic Impact and Healthiness Scores}
Nutrient profiling of food products is commonly used to provide an easy-to-understand recommendation for healthy eating, typically in a form of single numerical or graded scores. The ubiquity of such measures raises an interesting question whether or not diabetics and pre-diabetics can just rely on the dietary healthiness scores to make informed food choices for glycemic control. 

\textbf{Measuring Healthiness Scores}
The Food Standard Agency (FSA) of the UK published a guideline to measure the healthiness of food based on the amounts of fat, saturated fat, sugar, and salt per 100g/ml for a product\footnote{\url{https://www.food.gov.uk/sites/default/files/media/document/fop-guidance_0.pdf}}. 
Red, amber, and green color coding is used to represent the healthiness of each nutrient.
Following the procedures mentioned in Rokicki et al. \cite{Rokicki2018}\footnotemark, we computed the FSA healthiness score by adding up the scores in each nutrient, resulting in a final healthiness score ranging from 4 to 12. T\textit{he higher the score, the unhealthier the recipe in general}. 

\addtocounter{footnote}{-1}
\stepcounter{footnote}\footnotetext{Allrecipes provides the amount sodium (Na) instead of salt (NaCl), so we multiply Na by 2.54 to derive the amount of NaCl.}

\textbf{Correlational Analysis} 
Our crowdsourced judgements of the recipes' glycemic impact is an ordinal value ranging from [1, 5] where 1 = unlikely to have low glycemic impact, whereas 5 = highly likely to have low glycemic impact. Therefore, \textit{the higher the score, the healthier the recipe for diabetics}.

We computed the Pearson correlation coefficient of the crowdsourced judgements and the FSA healthiness scores for 990 recipes and found significant correlations ($p<0.01$) across all pairwise comparisons. Overall, there is a small correlation ($r=0.189$) between the glycemic impact and the total healthiness scores of recipes. Naturally, the glycemic impact judgements have a weak negative correlation ($r=-0.28$) with the FSA scores for total sugars, i.e., some low glycemic impact recipes tend to have a healthier amount of sugars. Together, this suggests that diabetics and pre-diabetics cannot simply rely on a general measurement of dietary healthiness to effectively control their glycemic responses. Interestingly, we also observed small positive correlations between the glycemic impact judgements and the FSA scores for fat ($r=0.299$), saturated fat ($r=0.19$), and salt ($r=0.148$), indicating that some low glycemic impact recipes do not have a healthy amount of those nutrients. The finding emphasizes the complementary nature of the glycemic potency and the healthiness of recipes in helping diabetics and pre-diabetics make informed dietary decisions.

\textbf{Between-Recipe-Class Differences} 
Next, we investigated the differences in the FSA healthiness scores between the two classes of recipes. To that end, we computed  the Kruskal-Wallis H test with Dunn's test on the average fat, saturated fat, sugars, salt, and total FSA scores between the UD and HD recipes. We found significant between-group differences in the FSA scores for fat ($p<0.001$) and saturated fat ($p<0.05$), but not for sugar, salt and total healthiness. Consistent with the correlational analysis, UD recipes have a higher average FSA score for sugar than HD recipes, i.e., they contain a less healthy amount of total sugars. On the other hand, UD recipes tend to have the lower average FSA scores for other nutrients as well as the total healthiness. In other words, except for their low glycemic impact association, UD recipes can be less healthy than HD recipes overall. This affirms the fact that diabetics and pre-diabetics are likely to benefit the most from using multiple and complementary healthiness measures when making optimal food choices. Together, RQ1's findings help motivate the significance of the proposed UD recipe classification task.

\section{Recipe Classification}
\label{sec:methods}

In this section, we present variants of recipe classification models to identify unhealthy-for-diabetics (UD) recipes in Recipe990. Each variant is defined by a combination of the recipe representation method and the underlying classification algorithm. Two well-known classification algorithms are explored: Logistic Regression (\textbf{LR})\footnote{\url{https://scikit-learn.org/}} and LightGBM (\textbf{LGBM}) \cite{Ke2017}\footnote{\url{https://pypi.org/project/lightgbm/2.2.2/}}. Next, the following recipe representation methods are considered:

\textbf{Bag-of-Words (BoW) Models}:
We included three variants of BoW representations. First, \textbf{BoW-basic} uses single words (unigram) extracted from the Recipe990 dataset as features and feature count as feature values. Next, for \textbf{BoW-parsed}, we parsed and removed numerals, quantity words, and other comment texts in the ingredient section of the recipes following the procedures used in \cite{Salvador2017} before feature extraction. Lastly, \textbf{NB-BoW} uses the same unigram feature set as BoW-basic but utilizes Naive Bayes log-count ratios, popularized and shown to be highly effective in \cite{SidaWang2012}, as a weight to feature count.

\textbf{Embedding Models}: 
Five well-known word (\textbf{word2vec}\footnotemark\cite{Mikolov2013}, \textbf{GloVe}\footnotemark\cite{Pennington2014}, \textbf{fastText}\addtocounter{footnote}{-2}\footnotemark\addtocounter{footnote}{+1}\cite{Bojanowski2017}), sentence (\textbf{skip-thoughts}\footnotemark \cite{kiros2015skip}), and paragraph (\textbf{doc2vec}\addtocounter{footnote}{-4}\footnotemark \cite{Le2014}) embedding models are considered.
For each word embedding model, we included both the pre-trained embedding trained on standard corpora and the model trained on our Recipe54K dataset. In particular, the pre-trained models of word2vec, GloVe, and fastText are trained on Google news, Wikipedia 2014, and Wikipedia 2017, respectively\footnotemark. Lastly, we also evaluated another word2vec variant trained on the Recipe1M dataset \cite{Salvador2017}.

\textbf{Nutritional Properties}:
As non-textual features, we included 19 nutritional properties and dry weight, resulting in 20 nutritional features (\textbf{NU}).

From the original Recipe54K dataset, we performed additional data preprocessing steps before generating any recipe representations by (1) converting any words that appeared less than 5 times into an unknown token (UNK) and discarding them, and (2) removing all punctuation marks. After this, the vocabulary sizes for Recipe54K and Recipe990 were 6,716 and 2,992, respectively. Additionally, every feature in the aforementioned representations except NB-BoW is scaled to the mean of zero and variance of one.

\addtocounter{footnote}{-3}
\stepcounter{footnote}\footnotetext{https://www.pydoc.io/pypi/gensim-3.2.0/index.html}
\stepcounter{footnote}\footnotetext{https://github.com/maciejkula/glove-python}
\stepcounter{footnote}\footnotetext{https://github.com/RaRe-Technologies/gensim-data}
\stepcounter{footnote}\footnotetext{https://github.com/sanyam5/skip-thoughts}

\section{Experimental Protocols}
\label{sec:experiment}

\subsection{Model Evaluation}
\label{sec:data_split}
We employed nested cross-validation to train and evaluate the classification models. Specifically, we randomly split and stratified the data into 5 outer folds and 5 inner folds. For each outer fold, there are approximately 800 recipes in the training set and 200 recipes in the test set. Then, we performed inner 5-fold cross-validation using the training set in each outer fold for hyperparameter tuning and selecting the optimal classification probability thresholds. In this stage, each inner fold is composed of roughly 640 recipes for the training set and 160 recipes for the test set. To evaluate the effectiveness of different models, we employed standard performance metrics: recall, precision, and F1, commonly used in the evaluation of text classification.  

\subsection{Hyperparameter Tuning}
\label{sec:tuning}
\textbf{Representation models} We optimally set the number of dimensions for all word embeddings to 300. The hyperparameters of Recipe54k-trained models are tuned within the nested cross-validation. We fixed the sampling rate to $10^{-3}$ and search the window size ranging from 5 to 40, the learning rate ranging from 0.025 to 0.05, and the iteration number ranging from 25 to 100.

The sentence and paragraph embeddings are both initialized with Recipe1M-trained word2vec and further trained on Recipe54k. For the sentence embedding, we set the number of dimensions to 1200, learning rate to \num{3e-4}, batch size to 32, and optimally stopped at around 25,000 steps. Next, a distributed bag-of-words (DBOW) version of doc2vec is used for paragraph embedding. The hyperparameters are tuned with the same procedures as tuning the word embeddings except the sampling rate is set to $10^{-5}$.

\textbf{Classification algorithms} For LR, we used liblinear as the solver and searched for the optimal regularization parameter C ranging from 0.01 to 1000. We tuned ten hyperparameters for LGBM, such as the learning rate, the number of leaves, the max depth for tree model, etc.

\textbf{Probability thresholds} For each classification model, we selected the optimal classification probability threshold through a grid search ranging from 0.45 to 0.55.

\section{Results}
\label{sec:results}

\subsection*{RQ2: Effectiveness of Recipe Embeddings}

\begin{table*}
\caption{Performance of different classification models}
\label{tbl:performance1}
\centering
\scalebox{0.8}{
\begin{threeparttable}
\begin{tabular}{llrrrrrr}
\toprule
Category & & & LR & & & LGBM & \\
& Representation & F1 & precision & recall & F1 & precision & recall\\
\midrule
                           & BoW-basic              & 0.76  & 0.81  & 0.717 & 0.76  & 0.77  & 0.751 \\
Bag-of-words               & BoW-parsed             & 0.764 & 0.798 & 0.733 & 0.755* & 0.757 & 0.753 \\
                           & NB-BoW                 & \textit{0.817} & 0.844 & 0.794 & \textit{0.77}  & 0.774 & 0.767 \\
\midrule
                           & Pretrained             &       &       &       &       &       &       \\
\cmidrule{2-8}
                           & word2vec               & 0.779 & 0.793 & 0.771 & 0.774 & 0.77  & 0.779 \\
                           & glove                  & 0.784 & 0.789 & 0.784 & \textit{0.802} & 0.814 & 0.791 \\
                           & fastText               & 0.78  & 0.777 & 0.785 & 0.761 & 0.779 & 0.749 \\
\cmidrule{2-8}
Word embedding             & Recipe54k-trained      &       &       &       &       &       &       \\
\cmidrule{2-8}
                           & word2vec               & 0.78  & 0.787 & 0.778 & 0.779 & 0.78  & 0.781 \\
                           & glove                  & 0.783 & 0.783 & 0.788 & 0.783 & 0.793 & 0.775 \\
                           & fastText               & 0.787 & 0.778 & 0.8   & 0.755* & 0.76  & 0.753 \\
\cmidrule{2-8}
                           & Recipe1M-trained       &       &       &       &       &       &       \\
\cmidrule{2-8}
                           & word2vec               &\textit{0.815} & 0.815 & 0.816 & 0.787 & 0.823 & 0.759 \\
\midrule
Sentence embedding         & skip thoughts          & 0.79  & 0.803 & 0.779 & 0.79  & 0.818 & 0.767 \\
\midrule
Paragraph embedding        & doc2vec                & 0.745* & 0.761 & 0.731 & 0.763 & 0.78  & 0.749 \\
\midrule
                           & NU only                  & 0.825 & 0.83  & 0.822 & 0.811 & 0.81  & 0.812 \\
Nutritional properties (NU) & NU + NB-BoW              & \textit{0.85}  & 0.859 & 0.842 & 0.831 & 0.823 & 0.84  \\
                           & NU + best word embedding & 0.836 & 0.851 & 0.824 & \textbf{0.854} & 0.852 & 0.858 \\
                           & NU + sentence embedding  & 0.821 & 0.857 & 0.79  & 0.843 & 0.847 & 0.839 \\
\bottomrule
\end{tabular}
\begin{tablenotes}
\footnotesize
\item The best overall results are in bold. The best results for each category are in italics. Dunn's multiple comparison in the set of LR or LGBM following the Kruskal-Wallis test: Significance found in comparison to the best variant in LR or LGBM set (\textit{*p < .05})
\end{tablenotes}
\end{threeparttable}
}
\end{table*}

Table ~\ref{tbl:performance1} compares the effectiveness of different classification models on the Recipe990 dataset. We use the notation "\textit{representation model} + \textit{classification algorithm}", e.g., BoW+LR, when referring to a specific variant. To compare the performance of these variants, we computed Kruskal-Wallis H tests with Dunn's multiple comparison test for two sets of LR or LGBM variants in F1 and found significant differences ($p<0.01$) between both. Due to space constraint, we only reported the pairwise comparisons against the best variant in each set on the F1 metric at the significance level of 0.05.

\textbf{Bag-of-Words} 
First, NB-BoW outperforms BoW-basic and BoW-parsed, confirming its superior performance in text classification tasks. Interestingly, NB-BoW+LR is also the best variant overall in F1 (0.817) among all textual content-only based models. We did not find significant difference between BoW-basic and BoW-parsed. The removal of quantity-related words did not affect the UD recipe classification performance.

\textbf{Word Embeddings} 
Overall, there is no significant difference in the performance of various word embedding-based models.
Next, none of the word embedding variants outperform NB-BoW+LR though they perform generally better than BoW-basic and BoW-parsed. Within the word embedding models, the Recipe1M pre-trained models outperform the Recipe54K-trained models. Both models perform better than those pre-trained on a large general corpus. Compared to all the other word embedding variants, Recipe1M pre-trained word2vec+LR achieves the best F1 score (0.815), whereas the pre-trained Glove+LGBM variant achieves the best F1 score (0.802) among the LGBM variants.

\textbf{Sentence Embedding} 
Similar to the results of the word embeddings, no sentence embedding models outperform NB-BoW+LR. The two sentence embedding variants (skip-thoughts+LR and skip-thoughts+LGBM) tend to perform equally well as most word embedding variants. This may suggest that the effectiveness of the UD recipe classification models does not depend on a fine-grained encoding of sentence semantics, which is one of the advantages of sentence embeddings.

\textbf{Paragraph Embedding} 
All paragraph embedding variants perform the worst in all metrics. This was even after we tried several optimization techniques, e.g., initializing the weights of doc2vec with the pre-trained Recipe1M word2vec and following the suggestion in \cite{lau2016empirical}. Our results are consistent with prior NLP research which found doc2vec to be highly task-dependent and difficult to train and optimize \cite{lau2016empirical}.

\textbf{Nutritional Features}
Exclusively employing the nutritional features results in a higher F1 compared to BoW-NB. Total carbohydrates is overall most important feature (1.236) followed by protein (-0.676) and dry weight (0.43). Then, we selectively combined the best BoW, word embedding, and sentence embedding features to the nutritional features. We used Recipe1M-trained word2vec for LR variants and pre-trained glove for LGBM variants. Ultimately, the combined pre-trained glove and nutritional features achieves the best F1 score of 0.854.

\textbf{Analysis of Performance}
We ranked the NB-weights \cite{SidaWang2012} and found a few pasta-related words, e.g., linguine, al dente, fettucini, which occur more frequently in the UD recipes than in the HD recipes. Many words associated with high carbohydrates were also found, e.g., cake, chocolate and cookies. The HD recipes are associated with meat and a high proportion of protein, such as steaks, rub, and swordfish. Furthermore, our model was able to successfully identify the glycemic impact differences between recipes even if they have the same carbohydrate contents. For example, given comparable carbohydrate contents, the model would correctly identify "conrad's spaghetti and meat sauce" as UD recipe and "south western spaghetti squash" as HD recipe.

\textbf{Hard-to-Classify Recipes}
After reviewing the misclassified cases, we found that many ethnic recipes with cuisine-specific sauces as ingredients, such as African peanut soup, menudo rojo (Latin), tomato chutney (India), are more difficult for the model to classify. Because the dietary glycemic impact is dependent of the actual amount of foods consumed, it is difficult to determine the impact without knowing the actual consumption sizes of those sauces. It is also equally challenging for humans to judge whether the recipe is UD when they are not familiar with such ingredients. On the other hand, recipes with almost no carbohydrates are one of the easiest to classify. 

\section{Conclusion \& Limitations}
In this paper, we investigated the feasibility of a data-driven and crowdsourcing approach to estimate the glycemic impact of cooking recipes. We first obtained the human judgements of recipes' glycemic impact through online crowdsourcing and created a dataset of 990 labeled recipes (Recipe990). By computing the recipes' FSA healthinesss scores and analyzing their relationship with the crowdsourced glycemic impact ratings, we found that many recipes with high healthiness scores might not be suitable for diabetics to consume due to their high glycemic potency. On the other hand, some low glycemic impact recipes contained relatively unhealthy amounts of fat, saturated fat, and salt. Therefore, the healthiness score and the dietary glycemic impact can be seen as two complementary measures. Next, we formulated glycemic impact inference as a binary classification problem and investigated the differences in the effectiveness of twelve recipe representations on the classification task. We showed that the best model was highly effective in classifying unhealthy-for-diabetics recipes. The more sophisticated recipe representations were not as effective as the nutritional information.

This work is not without limitations. First, due to its small size (N = 990), the models trained on Recipe990 are prone to overfitting. Next, the crowdsourced labels and the prediction results have not been validated by nutritional experts, so the model may be of limited use in real-world clinical applications. Lastly, most of the models used in our study simply treat quantity-related words in the recipe as a unique word, ignoring the impact of their proximity to the ingredients.

\section{Acknowledgement}
This research is supported by the National Research Foundation, Prime Minister's Office, Singapore under its International Research Centres in Singapore Funding Initiative.

\balance{}

\bibliographystyle{ACM-Reference-Format}
\bibliography{main}

\end{document}